\pdfoutput=1
\documentclass{webofc}
\usepackage[varg]{txfonts}
\usepackage{hyperref}
\usepackage{url}
\usepackage{tikz}
\usepackage{wrapfig}
\usetikzlibrary{arrows.meta,decorations.pathmorphing,calc}

\newif\ifprelimlogo
\prelimlogotrue
\hypersetup{colorlinks=true,citecolor=blue,urlcolor=blue,linkcolor=blue}

\newcommand{\Kstar}{\ensuremath{K^{*}(892)}}
\newcommand{\Kst}{\ensuremath{K^{*}}}
\newcommand{\Lam}{\ensuremath{\Lambda}}
\newcommand{\Lbar}{\ensuremath{\bar{\Lambda}}}
\newcommand{\pbar}{\ensuremath{\bar{p}}}
\newcommand{\Egamma}{\ensuremath{E_{\gamma}}}
\newcommand{\GeV}{\ensuremath{\mathrm{GeV}}}
\newsavebox{\formtab}%

\begin{document}

\title{Strangeness Production at GlueX}

\author{\firstname{Hao} \lastname{Li}\inst{1}\fnsep\thanks{\email{haoli@wm.edu}}
        \ (for the GlueX Collaboration)
}

\institute{William \& Mary, Williamsburg, Virginia 23185, USA}

\abstract{%
The GlueX experiment at Jefferson Lab is well suited for exclusive
studies of strangeness production, combining a linearly polarized
photon beam at $8.0 < \Egamma < 8.8~\GeV$ with a large-acceptance
detector that reconstructs weakly decaying hyperons. We outline a
research program built on three aspects: (i)~\emph{spectroscopy} of
strange mesons, where many excited $K^{*}$ states remain
unestablished and crypto-exotic strange hybrids must be identified
from combined spectrum and decay information, anchored by the \Kstar\
as a standard candle;
(ii)~\emph{spin dynamics}, where the first precise measurements of the
complete spin-density matrix elements in the photoproduced
$K^{*}(892)\Lambda$ system beyond the Schilling formalism
constrain the $t$-channel strange production mechanism; and
(iii)~\emph{hadronization}, probed through the newly published
$p\pbar$, $\Lam\Lbar$, and $p\Lbar$ photoproduction cross sections
and emerging $\Lam\Lbar$ spin-correlation measurements. We place
these efforts in the context of the broader GlueX strangeness
program and give an outlook toward the high-statistics GlueX-II era.
}

\maketitle

\section{Introduction}
\label{sec:intro}

Strangeness production in polarized photon-induced reactions provides a unique
laboratory of non-perturbative QCD where spectroscopy, spin dynamics,
and hadronization can be studied simultaneously.
At GlueX energies, $s$-channel resonance production is negligible, $t$-channel exchange dominates, and the spin-dependent
observables map directly onto production amplitudes. This contribution
organizes the GlueX effort in strangeness production and measurement into a coherent research agenda
of three pillars:
\begin{enumerate}\setlength{\itemsep}{1pt}\setlength{\parskip}{0pt}
\item \textbf{Spectroscopy}: mapping the excited strange-meson
  spectrum ($K\pi$, $K\pi\pi$, \dots) and strange baryons
  through mass spectra and partial-wave analyses (Sect.~\ref{sec:spectroscopy});
\item \textbf{Spin dynamics}: determining production amplitudes
  of reactions with open strangeness such as $\vec{\gamma}p \to \Kst Y$ through complete
  spin-density matrix elements (SDMEs) and hyperon polarization
  observables (Sect.~\ref{sec:spin});
\item \textbf{Hadronization}: understanding the role of confined strange quarks in strange-hadron formation through cross sections, mass spectra, and spin correlations
  of exclusive photoproduced baryon--antibaryon systems (Sect.~\ref{sec:hadronization}).
\end{enumerate}
They serve the GlueX mission to map the spectrum of light hadrons and to understand the role of gluonic excitations through the search for hybrid mesons. For the spin-exotic $\pi_{1}(1600)$, GlueX has recently set
the first upper limits on photoproduction~\cite{GlueX:Pi1}.
In the strange sector, this program is comparatively underdeveloped:
many quark-model $K^{*}_{J}$ states remain unestablished
(Fig.~\ref{fig:kpi-spectrum}, middle).
Completing the strange spectrum, particularly in the axial and tensor
sectors, will provide important information for comparing flavor
partners and interpreting the quantum numbers of possible exotic
states in the non-strange sector. Because open-strange mesons lack a
definite charge-conjugation quantum number, low-lying strange hybrids
share spin and parity with conventional kaon
excitations~\cite{Shastry:2022plb,Dudek:2011hyb,Chen:2026k1690}. 
Strange hybrid identification therefore requires combined evidence from the mass spectrum and spin-dependent decay dynamics. More specifically, relevant signatures may include anomalies in the spectrum, such as supernumerary states, characteristic channel couplings, and possible mixing with conventional mesons.

\section{Strange meson spectroscopy at GlueX}
\label{sec:gluex}
\label{sec:spectroscopy}

The GlueX experiment in Hall~D at Jefferson Lab is building the world's
largest photoproduction data set at $6 < \Egamma < 11.4~\GeV$. A
linearly polarized photon beam, produced by coherent bremsstrahlung
with polarization up to ${\sim}40\%$ in the coherent peak
($8.0<\Egamma<8.8~\GeV$), is incident on a liquid-hydrogen target
inside a near-hermetic spectrometer built around a 2~T solenoid, with
tracking and calorimetry covering central and forward
regions~\cite{GlueX:NIM}.

\begin{figure}[!h]
\centering
\begin{minipage}[c]{0.72\textwidth}
\centering
\begin{tikzpicture}
\node[anchor=south west, inner sep=0] (img)
  {\includegraphics[width=\linewidth,clip]{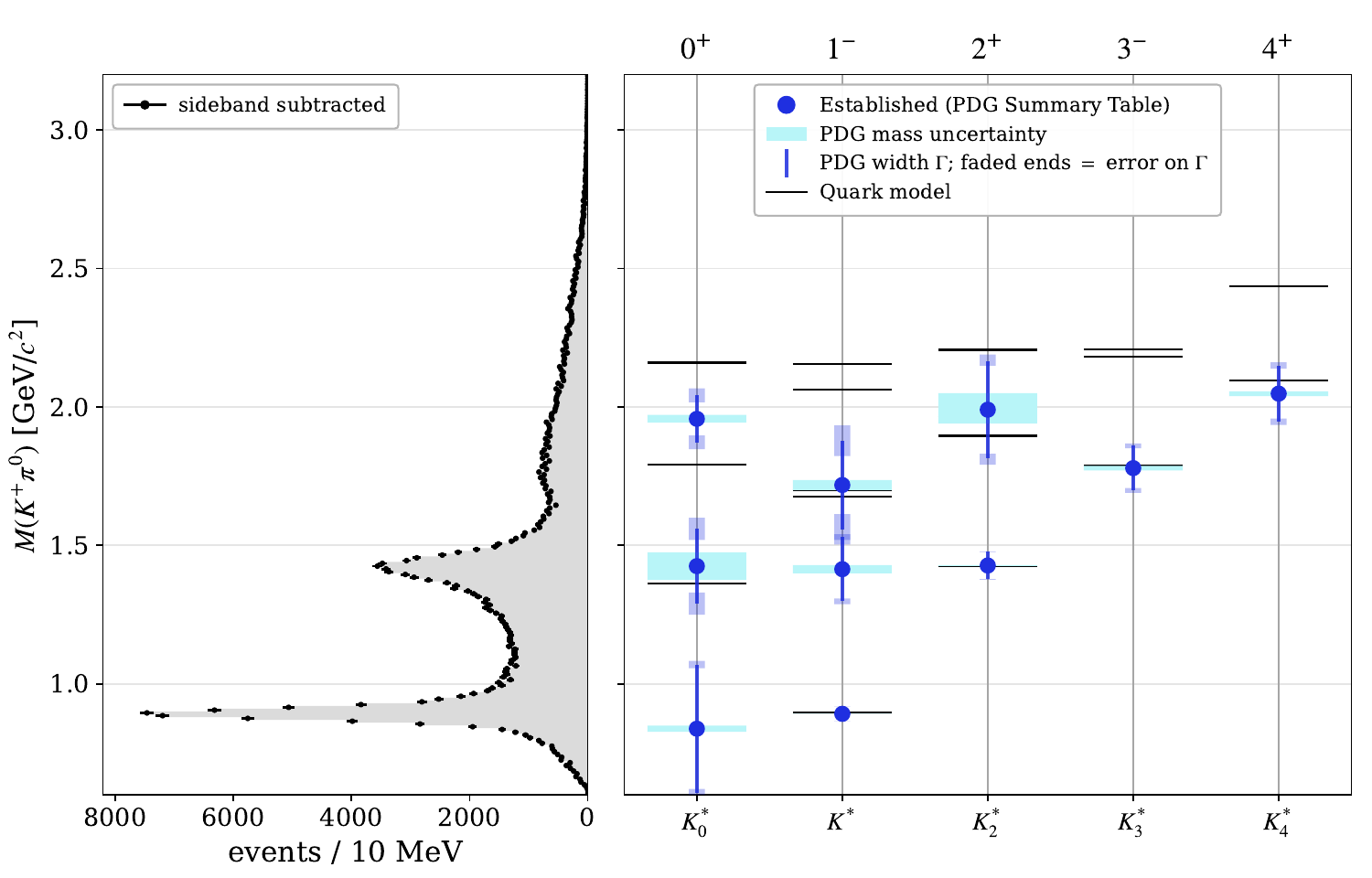}};
\ifprelimlogo
\path (img.south west) -- (img.south east) coordinate[pos=0.36] (wx);
\path (img.south west) -- (img.north west) coordinate[pos=0.72] (wy);
\node[opacity=0.55] at (wx |- wy)
  {\includegraphics[width=0.12\linewidth]{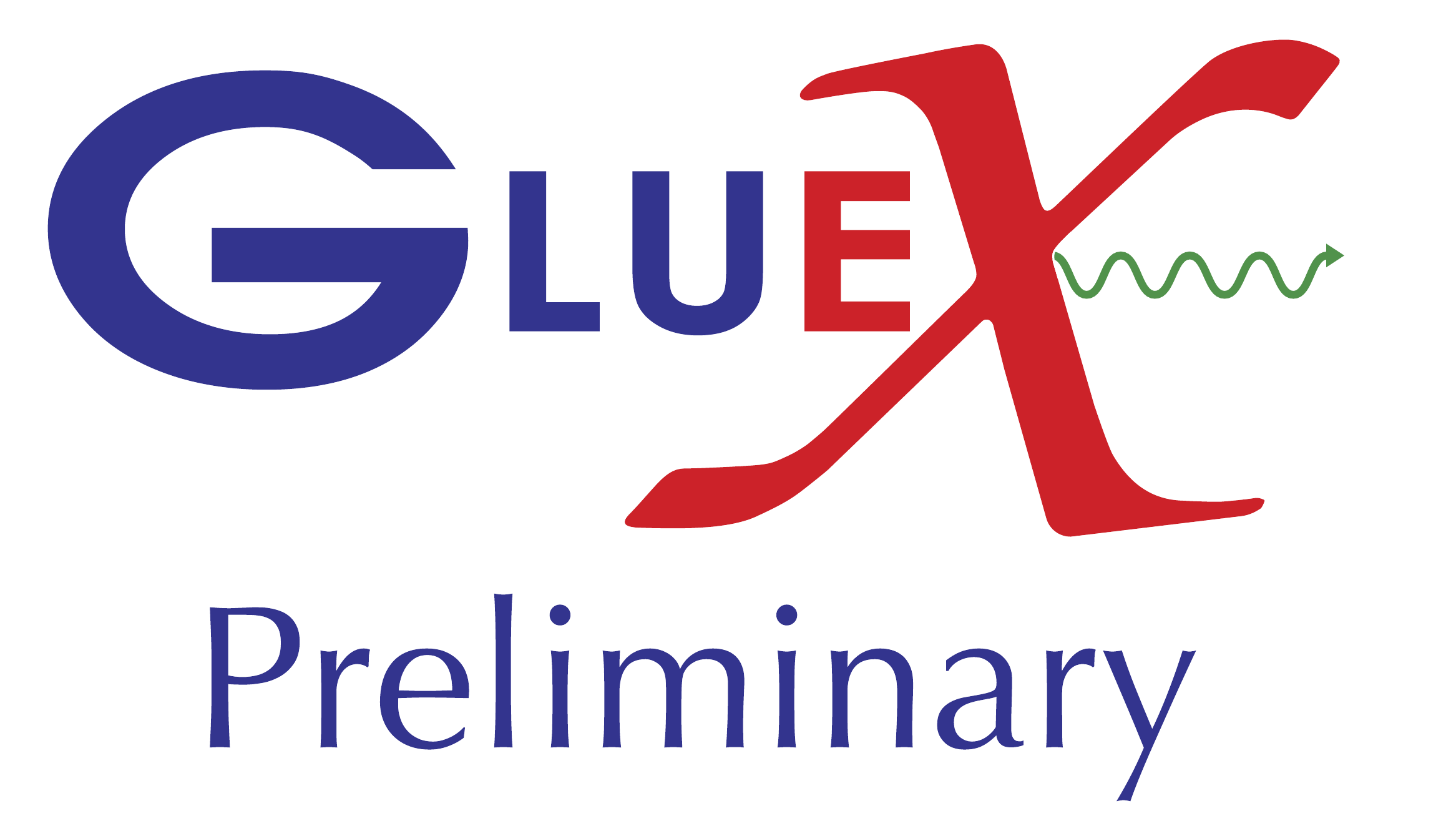}};
\fi
\end{tikzpicture}
\end{minipage}\hspace{0.012\textwidth}%
\begin{minipage}[c]{0.245\textwidth}
\centering
\includegraphics[width=\linewidth,clip]{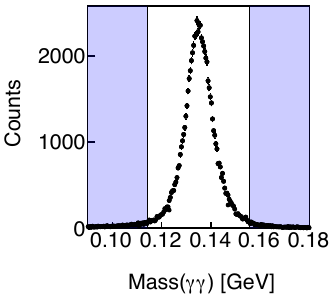}\\[3pt]
\includegraphics[width=\linewidth,clip]{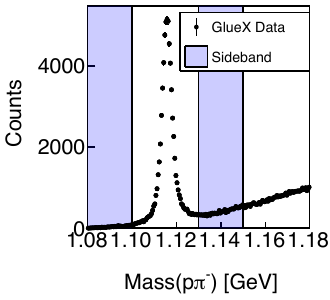}
\end{minipage}
\caption{Left: sideband-subtracted $K^{+}\pi^{0}$ invariant-mass
  spectrum from GlueX data for $\gamma p \to K^{+}\pi^{0}\Lam$,
  dominated by the \Kstar\ with clear structure in the
  $K^{*}_{0}/K^{*}_{2}$ region near $1.43~\GeV$. Middle: the excited
  $K^{*}$ spectrum of established states (PDG)~\cite{PDG:2024} compared with quark-model
  predictions~\cite{Ebert:2009ub}, aligned on the same mass axis; the
  GlueX $K\pi$ data span the mass range where several predicted states
  remain unestablished. Right: the $\gamma\gamma$ (top) and $p\pi^{-}$
  (bottom) invariant-mass distributions with the $\pi^{0}$ and \Lam\
  signal and sideband regions used for the subtraction.}
\label{fig:kpi-spectrum}
\end{figure}

GlueX is particularly well suited for
strangeness measurements:
(i)~the linear beam polarization acts as a filter on the naturality of
the $t$-channel exchange and provides access for amplitude analysis;
(ii)~exclusive event reconstruction with charged tracks and neutral
showers fully constrains the final state; and
(iii)~multi-stage reconstruction of detached vertices identifies weakly
decaying hyperons ($\Lam \to p\pi^{-}$, $\Lbar \to \pbar\pi^{+}$) and
turns their self-analyzing decays into
polarimeters~\cite{Lee:1957qs}.

Figure~\ref{fig:kpi-spectrum} shows the GlueX $K^{+}\pi^{0}$
invariant-mass spectrum from the exclusive reaction
$\gamma p \to K^{+}\pi^{0}\Lam$ alongside the known and predicted
excited $K^{*}$ spectrum: beyond the dominant \Kstar, the data show a
pronounced structure near $1.43~\GeV$ compatible with overlapping
$K^{*}_{0}(1430)/K^{*}_{2}(1430)$ resonances, and significant yield
extending above $1.6~\GeV$. A clean sideband-subtracted selection
controls the backgrounds underneath the $\pi^{0}$ and \Lam\ peaks in the
$\gamma\gamma$ and $p\pi^{-}$ mass distributions
(Fig.~\ref{fig:kpi-spectrum}, right).

The GlueX strangeness program proceeds in a three-step roadmap:
(1)~establish the \Kstar\ as a ``standard candle'' state, benchmarking precision measurements of the spin observables (Sect.~\ref{sec:spin}) and the comparison between charged ($\Kst{}^{+}\Lam$, $\Kst{}^{+}\Sigma^{0}$) and neutral ($\Kst{}^{0}\Sigma^{+}$) exchange in channels that share the same formalism;
(2)~perform amplitude analyses of the excited $K^{*}$ region at higher
$K\pi$ mass, resolving the interference among $S$-, $P$-, and
$D$-waves, where the weakly decaying hyperon polarization may provide
extra constraints on partial-wave
ambiguities~\cite{Guo:2025amb};
(3)~extend to $K\pi\pi$ and other multi-body strange systems, the
natural home of the strange partners of the axial and hybrid
candidates. The unprecedented statistics of the GlueX photoproduction data make the $K^{+}\pi^{0}\Lam$ channel an ideal testbed for methodology development within the GlueX
amplitude-analysis framework~\cite{GlueX:PWA}.

\section{Spin dynamics and production mechanisms}
\label{sec:spin}

For $\vec{\gamma} N \to \Kst\Lam$ with an unpolarized target proton, the Hilbert space $\mathbf{H}_{\gamma}\otimes\mathbf{H}_{K^{*}}\otimes\mathbf{H}_{\Lam}$ contains $2\times3\times2 = 12$ helicity states, and the joint system is described by a $12\times12$ spin-density matrix built from bilinears $A_{i}A_{j}^{*}$ of helicity amplitudes~\cite{Pichowsky:1996tn}. 
With the helicity labels of the $t$-channel exchange
diagram below and $\rho^{\gamma}(\Phi)$ the density matrix of a beam
linearly polarized at angle $\Phi$, the complete meson--hyperon spin-density matrix is:
\par\noindent
\begin{minipage}[c]{0.26\linewidth}
\centering
\begin{tikzpicture}[scale=0.79, every node/.style={transform shape},
    >=Latex,
    particle/.style={align=center, inner sep=1pt, font=\footnotesize},
    decayp/.style={inner sep=0.8pt, font=\footnotesize},
    inline/.style={inner sep=1pt, font=\scriptsize},
    vertex/.style={circle, fill=black, inner sep=1.1pt},
    dvertex/.style={circle, fill=black, inner sep=0.9pt},
    photon/.style={decorate, decoration={snake, amplitude=0.8pt, segment length=4pt}},
    vector/.style={decorate, decoration={snake, amplitude=0.55pt, segment length=3.2pt}},
    fermion/.style={->},
    exchange/.style={dashed}]
    \node[particle] (gamma) at (-1.60,1.05) {$\vec\gamma$\\$(\mathbf k,\lambda_\gamma)$};
    \node[particle] (nucleon) at (-1.60,-1.05) {$N$\\$(\mathbf p_N,\lambda_N)$};
    \node[vertex] (vtop) at (0,0.60) {};
    \node[vertex] (vbot) at (0,-0.60) {};
    \draw[photon, ->] (gamma) -- (vtop);
    \draw[fermion] (nucleon) -- (vbot);
    \draw[exchange] (vtop) -- node[inline, right=1pt] {$\kappa/K/K^{*}$} (vbot);
    \node[dvertex] (dV) at (1.10,1.05) {};
    \draw[vector, ->] (vtop) -- (dV);
    \node[inline] at (0.35,1.24) {$K^{*}(\mathbf q,\lambda_V)$};
    \draw[fermion] (dV) -- ++(0.58,0.34) node[decayp, right=-1pt] {$K$};
    \draw[fermion] (dV) -- ++(0.64,-0.16) node[decayp, right=-1pt] {$\pi$};
    \node[inline] at (1.34,0.72) {$\Omega_{K^{*}}$};
    \node[dvertex] (dY) at (1.10,-1.05) {};
    \draw[fermion] (vbot) -- (dY);
    \node[inline] at (0.35,-1.24) {$\Lambda(\mathbf p_Y,\lambda_Y)$};
    \draw[fermion] (dY) -- ++(0.64,0.16) node[decayp, right=-1pt] {$p$};
    \draw[fermion] (dY) -- ++(0.58,-0.34) node[decayp, right=-1pt] {$\pi^{-}$};
    \node[inline] at (1.34,-0.72) {$\Omega_{Y}$};
\end{tikzpicture}
\end{minipage}%
\hfill
\begin{minipage}[c]{0.72\linewidth}
{\small
\begin{equation}
\label{eq:double-sdme}
  \rho_{\lambda_{V}\lambda_{Y};\,\lambda_{V}'\lambda_{Y}'}(\Phi)
  \propto
    \sum_{\lambda_{\gamma}\lambda_{\gamma}'\lambda_{N}}
    A_{\lambda_{V}\lambda_{Y};\lambda_{\gamma}\lambda_{N}}\,
    \rho^{\gamma}(\Phi)_{\lambda_{\gamma}\lambda_{\gamma}'}\,
    A^{*}_{\lambda_{V}'\lambda_{Y}';\lambda_{\gamma}'\lambda_{N}}.
\end{equation}%
}%
\end{minipage}
\par\noindent
After enforcing normalization, hermiticity, and parity, different formalisms arise from distinct experimentally accessible projections of the complete spin-density matrix (Table~\ref{tab:formalisms}):
(a) the classic Schilling SDMEs probe only the $\Kst$ sector, $\rho^{K^{*}}_{\lambda_{V}\lambda_{V}'}(\Phi) = \sum_{\lambda_{Y}} \rho_{\lambda_{V}\lambda_{Y};\lambda_{V}'\lambda_{Y}}$, accessed through the $\Kst$ decay angles $\Omega_{K^{*}}$; (b) the \Lam\ polarization observables ($P$, $\Sigma$, $T$, $O_{x}$, $O_{z}$) characterize only the hyperon sector, $\rho^{\Lam}_{\lambda_{Y}\lambda_{Y}'}(\Phi) = \sum_{\lambda_{V}} \rho_{\lambda_{V}\lambda_{Y};\lambda_{V}\lambda_{Y}'}$, accessed through the \Lam\ decay angles $\Omega_{Y}$; and (c) the untraced spin-correlation sector of the double SDMEs retains the $\Kst\otimes\Lam$ spin information absent from both reduced formalisms, accessible only through the joint $(\Omega_{K^{*}},\Omega_{Y})$ distribution. 
In particular, the spin-correlation sector measures how the $\Kst$ spin alignment is correlated with
the recoil hyperon polarization, probing amplitude interferences
that reveal how strange-quark production and spin dynamics are encoded in the underlying helicity structure. This program is developed in collaboration with JPAC, aiming to map the now experimentally accessible meson--baryon joint spin-density matrix~\cite{Mathieu:2026meson} back to amplitude models and resolve production dynamics that remain hidden in reduced spin-density-matrix formalisms, with minimal model dependence.

\begin{table}[h] \vspace{3pt} \centering
\makeatletter
\renewcommand\@CaptionFont{\normalfont\footnotesize}%
\renewcommand\@makecaption[2]{\par\addvspace\abovecaptionskip
  \begingroup\@CaptionFont{\@CaptionTitleFont #1\@FloatCounterEnd\@FloatCounterSep}%
  \ignorespaces#2\par\endgroup\par\addvspace\belowcaptionskip}%
\makeatother
\caption{Decomposition of the complete meson--hyperon spin-density matrix for $\vec{\gamma} N \to \Kst\Lam$ into experimentally accessible formalisms (sub-sectors of the complete matrix indented; the Schilling and spin-correlation sectors partition the complete matrix, while the hyperon-polarization sector is a projection overlapping both).}
\label{tab:formalisms}
\vspace{2pt}
\begin{minipage}[c]{\linewidth}
\centering\footnotesize
\setlength{\tabcolsep}{3pt}%
\newcommand{\subrow}[1]{{\footnotesize #1}}
\sbox{\formtab}{%
\begin{tabular*}{\linewidth}{@{\extracolsep{\fill}}ll}
\hline
Formalism & Observable angular modulations$^{*}$ \\
\hline
Complete meson--hyperon spin-density matrix & 35 \\
\subrow{\quad Schilling (reduced $\Kst$ sector)~\cite{Schilling:1969um}} & \subrow{9} (Fig.~\ref{fig:sdme}) \\
\subrow{\quad Spin-correlation sector ($\Kst\otimes\Lam$)~\cite{Pichowsky:1996tn}} & \subrow{26} \\
\subrow{\quad Hyperon polarization (reduced \Lam   ~sector)~\cite{Ireland:2019uja}} & \subrow{5$^{\dagger}$} \\
\hline
\end{tabular*}}%
\usebox{\formtab}
\par\vspace{2pt}
\parbox{\wd\formtab}{\scriptsize\raggedleft
$^{*}$Modulations visible for a linearly polarized beam
and an unpolarized target, after applying normalization.\par
$^{\dagger}$Two representative hyperon observables ($\Sigma, P$), projected from the complete spin-density-matrix formalism fit, are shown in Fig.~\ref{fig:sigmaP}.\par}
\end{minipage}
\vspace{3pt}
\end{table}

\begin{figure}[!htb]
\begin{minipage}[t]{0.61\linewidth}
\centering
\begin{tikzpicture}
\node[anchor=south west, inner sep=0] (imgA)
  {\includegraphics[width=\linewidth,clip]{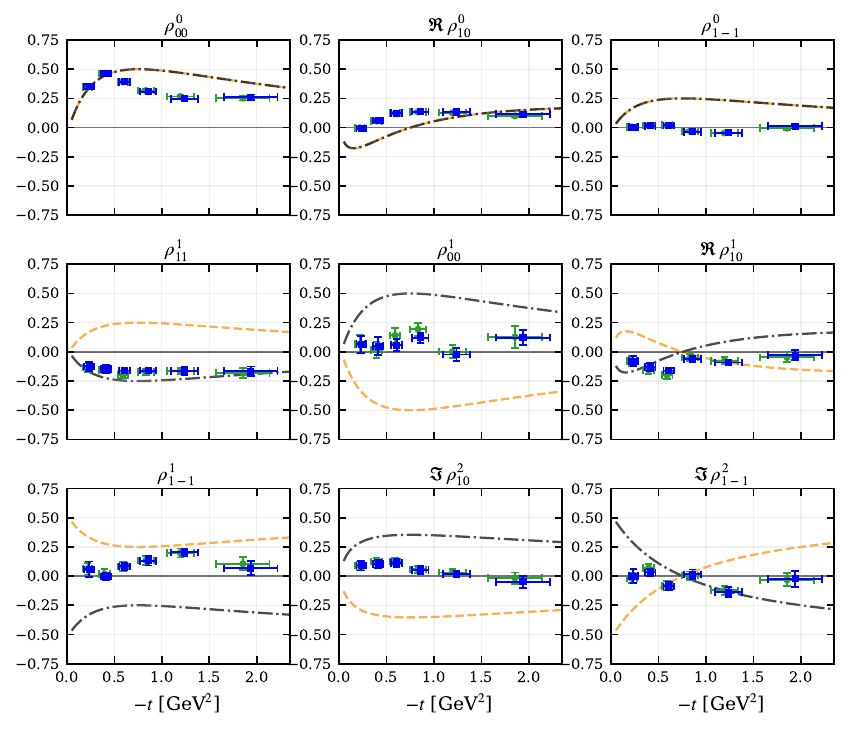}};
\ifprelimlogo
\path (imgA.south west) -- (imgA.south east) coordinate[pos=0.146] (ax);
\path (imgA.south west) -- (imgA.north west) coordinate[pos=0.759] (ay);
\node[opacity=0.55] at (ax |- ay)
  {\includegraphics[width=0.10\linewidth]{figures/gluex_preliminary}};
\fi
\end{tikzpicture}
\caption{Preliminary \Kstar\ spin-density matrix elements (helicity
  frame) in $\vec{\gamma} p \to \Kst{}^{+}\Lam$ versus momentum
  transfer $-t$: the Schilling formalism (green, statistical
  uncertainties) and the complete spin-density matrix formalism (blue,
  bootstrap uncertainties) are consistent. Lines show the expectations
  for a pure natural $\kappa$ (orange, dashed) or unnatural $K$
  (black, dash-dotted) exchange component.}
\label{fig:sdme}
\end{minipage}\hfill
\begin{minipage}[t]{0.37\linewidth}
\centering
\begin{tikzpicture}
\node[anchor=south west, inner sep=0] (imgB)
  {\includegraphics[width=\linewidth,clip]{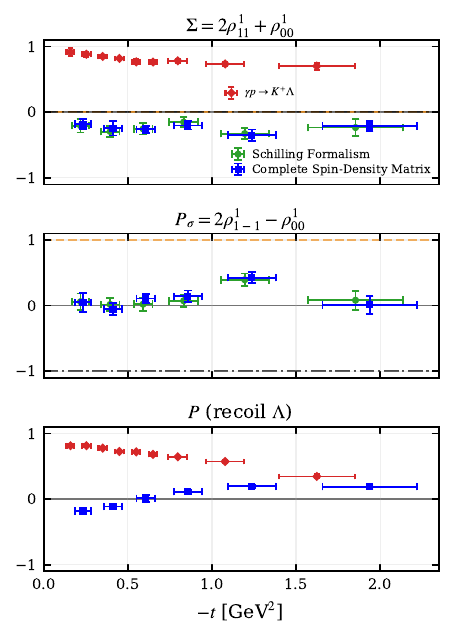}};
\ifprelimlogo
\path (imgB.south west) -- (imgB.south east) coordinate[pos=0.72] (bx);
\path (imgB.south west) -- (imgB.north west) coordinate[pos=0.175] (by);
\node[opacity=0.55] at (bx |- by)
  {\includegraphics[width=0.185\linewidth]{figures/gluex_preliminary}};
\fi
\end{tikzpicture}
\caption{Preliminary beam asymmetry $\Sigma$ (top), parity asymmetry
  $P_{\sigma}$ (middle), and recoil-$\Lam$ polarization $P$ (bottom)
  versus $-t$ for $\vec{\gamma}p \to \Kst{}^{+}\Lam$ (fits and
  colors as in Fig.~\ref{fig:sdme}), compared with the
  $\gamma p \to K^{+}\Lam$ channel~\cite{Furletov:2025kpl} (red).}
\label{fig:sigmaP}
\end{minipage}
\end{figure}

Figure~\ref{fig:sdme} shows the \Kstar\ SDMEs from the first fit of the complete $\Kst\Lam$ spin-density matrix beyond the Schilling formalism, compared with the Schilling-formalism fit; both are extracted in the helicity frame over a large range of momentum transfer $t$ with unbinned, extended maximum-likelihood fits. The
fits
use the \Kstar\ peak region, $0.84 < M(K^{+}\pi^{0}) < 0.94~\GeV$,
where a nonresonant $S$-wave contribution may be present
beneath the \Kstar. 
The consistency between the Schilling and complete-spin-density-matrix fits establishes the \Kstar\ as a benchmark for validating the analysis framework before its application to future study of excited $K^{*}$ states.
As easily interpretable baselines, Figs.~\ref{fig:sdme}
and~\ref{fig:sigmaP} also show the expectations for a pure natural
(scalar $\kappa$) or unnatural (pseudoscalar $K$) exchange
component~\cite{Ozaki:2009mj,Yu:2016pio,Wang:2019mid}; neither alone describes the
data over the full $t$ range. 
In addition, pure spin-0 exchange gives $\Sigma=0$
and $P_{\sigma}=\pm1$ exactly and constrains only the $\Kst$
sector, so no baseline is drawn for $P$.

The comparison to single-pseudoscalar photoproduction, $\gamma p \to K^{+}\Lam$, exposes the $t$-channel dynamics of strangeness exchange.
As shown in Fig.~\ref{fig:sigmaP}, the beam asymmetry for
$K^{+}\Lam$~\cite{Furletov:2025kpl} production is close to $+1$, indicating dominance of natural
$K^{*}$ exchange, whereas $\Kst\Lam$ shows a markedly different
pattern, reflecting a mix of natural and unnatural exchanges
($\kappa/K/K^{*}$) and a richer spin structure. 

These measurements
extend GlueX's earlier spin-dynamics results in the strange sector,
the beam asymmetry in
$\Sigma^{0}$ photoproduction~\cite{GlueX:Sigma0} and the
$\Lam(1520)$ spin-density matrix elements~\cite{GlueX:Lambda1520},
which likewise constrain $t$-channel exchange in the hyperon sector.
Quantitative predictions for these spin-dependent observables are currently tuned to lower beam energies dominated by $s$-channel resonance production~\cite{Yu:2016pio,Wang:2019mid,Wei:2020fmh}. Extending these calculations to GlueX energies, where $t$-channel exchange dominates, could provide valuable guidance for interpreting the measured spin dynamics.

\section{Strangeness in hadronization}
\label{sec:hadronization}

\begin{figure}[!htb]
\centering
\includegraphics[width=0.995\linewidth,trim=2 2 3 13,clip]{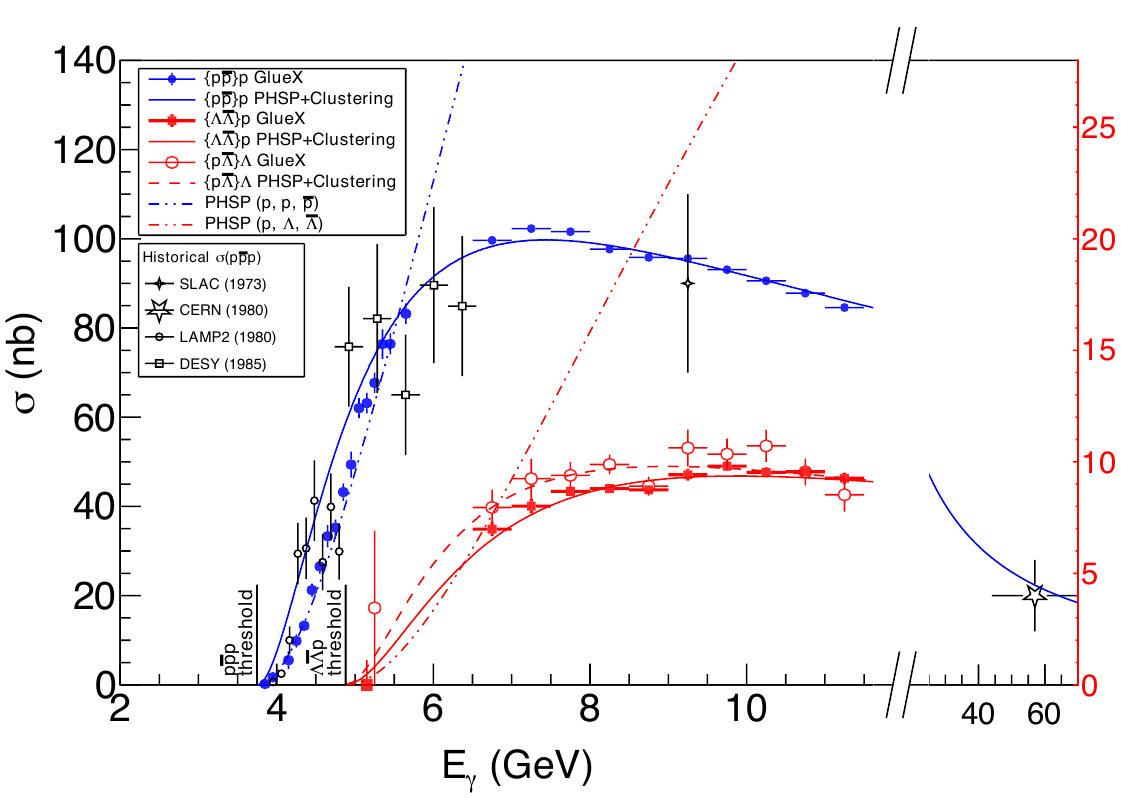}
\caption{Total cross sections for exclusive $\gamma p \to p\pbar
  p$, $\Lam\Lbar p$, and $p\Lbar\Lam$ photoproduction measured by
  GlueX~\cite{GlueX:BBbar}, compared with historical $p\pbar$ data and
  phase-space--based models.}
\label{fig:bbbar-xsec}
\end{figure}

\begin{wrapfigure}{r}{0.40\linewidth}
\centering
\includegraphics[width=\linewidth,trim=2 2 2 2,clip]{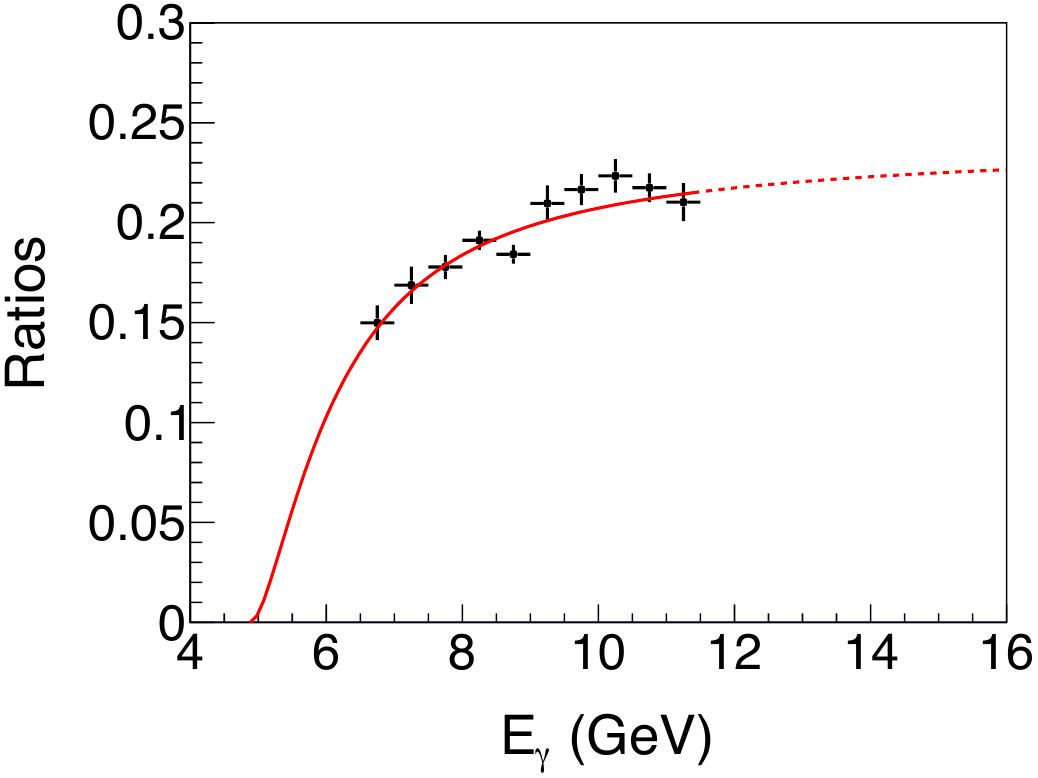}
\caption{Strange-to-non-strange cross section ratio versus beam
  energy~\cite{GlueX:BBbar}, related to the relative probability
  $P(s\bar{s})/P(u\bar{u})$ in hadronization.}
\label{fig:bbbar-ratio}
\end{wrapfigure}
The third pillar asks how strange hadrons emerge from confined quarks. Hadronization, the process by which quarks reorganize into color-neutral hadrons, remains one of the least constrained aspects of QCD. Exclusive baryon--antibaryon photoproduction probes this process in a controlled few-body system near the baryon--antibaryon mass threshold.
GlueX has
recently published the first measurement~\cite{GlueX:BBbar} of exclusive baryon--antibaryon
photoproduction cross sections combining $N\bar{N}$, $Y\bar{Y}$, and
$N\bar{Y}$ final states over a broad kinematic
range: $\gamma p \to \{p\pbar\} p$, $\{\Lam\Lbar\} p$, and
$\{\Lbar p\}\Lam$ (Fig.~\ref{fig:bbbar-xsec}). The reaction mechanism is
well described in terms of Regge-like single and double $t$-channel
exchange topologies, separated experimentally through the angular and momentum
distributions of the three final-state particles. The ratio of strange
to non-strange cross sections rises from threshold to a plateau of
${\sim}0.23$ (Fig.~\ref{fig:bbbar-ratio}), providing a direct handle
on the strangeness-suppression
factor of QCD hadronization in an exclusive, low-multiplicity
environment.

Two aspects extend this program beyond cross sections. First, the
invariant-mass spectra of the baryon--antibaryon pairs show a broad
threshold enhancement in the $p\pbar$, $\Lam\Lbar$, and $p\Lbar$
systems~\cite{GlueX:BBbar}, with no evidence for the $X(1835)$-related $p\pbar$ threshold structure reported in $J/\psi$ decays~\cite{BES:X1835};
final-state-interaction~\cite{Kerbikov:2004tj, Haidenbauer:2023zcu,
  Haidenbauer:2024smo} and sub-threshold
scalar-meson~\cite{Gutsche:2017xtm} interpretations are under study. Second, the parity-violating decays of \Lam\ and \Lbar\
reveal the first measurable $\Lam\Lbar$ spin-correlation pattern in the photoproduced baryon--antibaryon system~\cite{Li:2023thesis}, with precise interpretation ongoing with theoretical development. These correlations are sensitive to how quark
pairs reorganize and hadronize into correlated baryonic matter in
non-perturbative QCD. With them, the GlueX strangeness program can be
expanded toward the frontier of baryon-pair entanglement studies recently highlighted by BESIII and STAR~\cite{BES:Entangle, STAR:2026}.

\section{The broader GlueX strangeness program}
\label{sec:program}

\begin{figure}[!htb]
\centering
\begin{tikzpicture}
\node[anchor=south west, inner sep=0] (imgC)
  {\includegraphics[width=0.38\linewidth,clip]{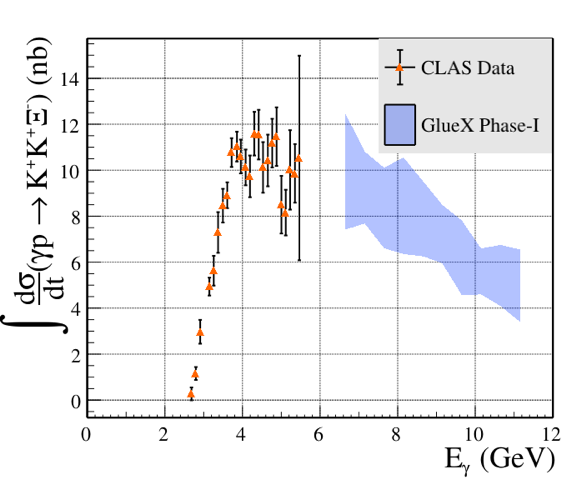}};
\ifprelimlogo
\path (imgC.south west) -- (imgC.south east) coordinate[pos=0.36] (cx);
\path (imgC.south west) -- (imgC.north west) coordinate[pos=0.84] (cy);
\node[opacity=0.55] at (cx |- cy)
  {\includegraphics[width=0.098\linewidth]{figures/gluex_preliminary}};
\fi
\end{tikzpicture}\hspace{0.03\linewidth}%
\begin{tikzpicture}
\node[anchor=south west, inner sep=0] (imgD)
  {\includegraphics[width=0.45\linewidth,clip]{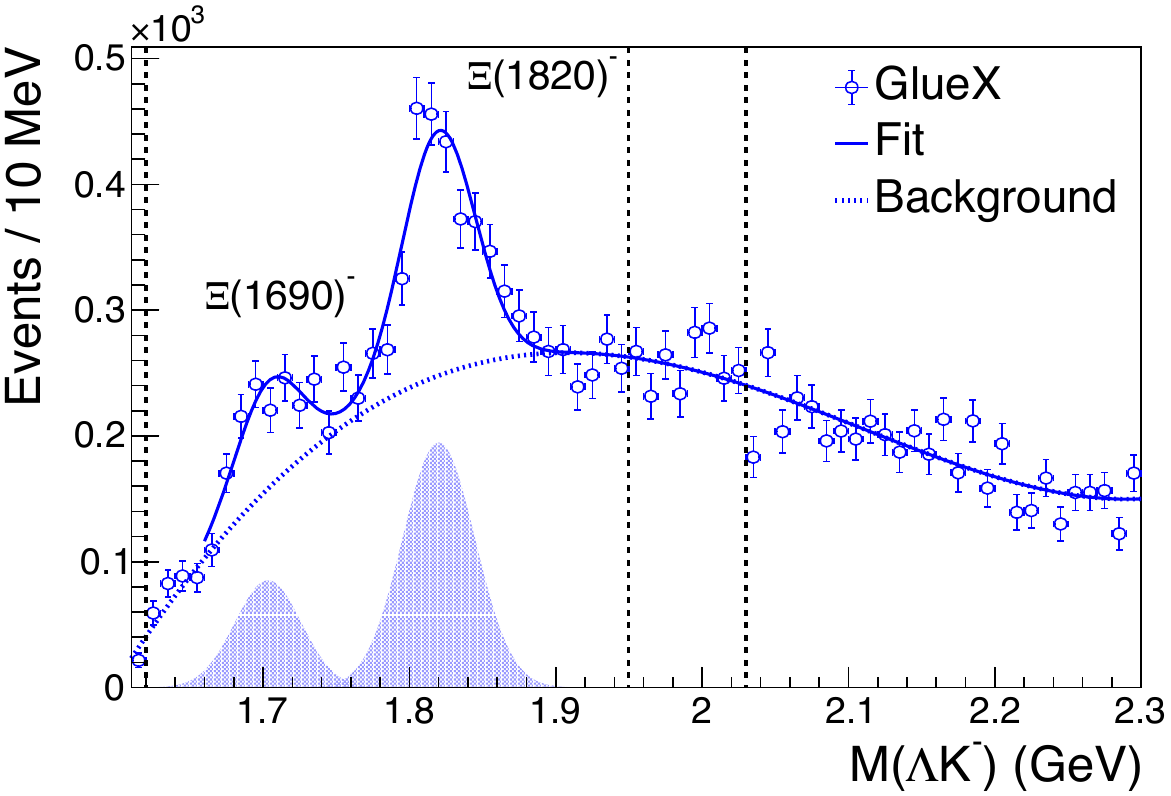}};
\ifprelimlogo
\path (imgD.south west) -- (imgD.south east) coordinate[pos=0.80] (dx);
\path (imgD.south west) -- (imgD.north west) coordinate[pos=0.25] (dy);
\node[opacity=0.55] at (dx |- dy)
  {\includegraphics[width=0.132\linewidth]{figures/gluex_preliminary}};
\fi
\end{tikzpicture}
\caption{Left: integrated $\Xi^{-}(1320)$ photoproduction cross
  section, comparing CLAS data with the GlueX Phase-I
  measurement~\cite{Hernandez:2025thesis} (band).
  Right: $\Lam K^{-}$ invariant-mass spectrum with the fit (solid) and
  its background component (dotted), showing $\Xi^{-}(1690)$ and the
  first $\Xi^{-}(1820) \to \Lam K^{-}$ signal in photoproduction.
  The GlueX results in both panels are preliminary; see also the
  review of Ref.~\cite{Crede:Cascades}.}
\label{fig:cascades}
\end{figure}

The three aspects elaborated above sit within a wider set of GlueX strangeness
results, surveyed earlier in Ref.~\cite{Pauli:HYP2022}. Ongoing
analyses address the $\Lam(1405)$
lineshape~\cite{Wickramaarachchi:2022mhi}, a long-standing benchmark for
chiral-unitary dynamics, and cascade spectroscopy
(Fig.~\ref{fig:cascades}): GlueX has measured $\Xi^{-}(1320)$ and
$\Xi^{-}(1530)$ photoproduction cross
sections~\cite{Hernandez:2025thesis} and observes the first $\Xi^{-}(1820) \to \Lam K^{-}$ signal in photoproduction, with many
quark-model $\Xi^{(*)}$ states still awaiting
confirmation~\cite{Crede:Cascades}. 
Future GlueX measurements in both meson--hyperon and hyperon--antihyperon channels can build on the amplitude-level methodologies developed and tested in the current broad GlueX strangeness program.

\section{Summary and outlook}
\label{sec:outlook}

The GlueX strangeness program is anchored by the precise measurement of the complete spin-density
matrix of the $\Kst\Lam$ system, validating an amplitude-analysis framework that will be carried into the planned analyses of the excited strange-meson spectrum. Exclusive baryon--antibaryon photoproduction
provides rich data on the role of strangeness in hadronization, from
cross-section ratios to threshold enhancements and spin correlations,
while cascade cross sections deliver new inputs to the PDG. The
forthcoming GlueX-II data set will significantly expand all three
aspects (spectroscopy, spin dynamics, and hadronization) toward a
comprehensive, amplitude-level picture of strange hadron production.

This research is supported by the Office of Nuclear Physics within the
U.S. Department of Energy Office of Science, under Award
No.~DE-SC0023978. We thank V.~Mathieu (JPAC), A.-C.~Wang, and N.-C.~Wei for valuable theoretical input and discussions. Claude Code (Anthropic, Claude Fable 5.1 model) was used to improve language and clarity; all AI-assisted text was reviewed by the author.
GlueX acknowledges the support of several funding
agencies and computing facilities (\url{http://gluex.org/thanks}).

\renewcommand{\bibfont}{\normalfont\fontsize{7}{7.6}\selectfont}
\setlength{\bibsep}{0pt}

\end{document}